\documentclass[a4paper,french]{rnti}
\usepackage{longtable}
\usepackage[ruled,vlined,french,titlenumbered]{algorithm2e}
\usepackage{amssymb}
\usepackage{amsmath}
\usepackage[dvips]{graphicx}
\newtheorem{definition}{Définition}
\newtheorem{proposition}{Proposition}

\newenvironment{preuve}[1][Preuve]{\noindent\textit{\textbf{#1. \ }}}{}
\newtheorem{exemple}{Exemple}

\titre{Nouvelle représentation concise exacte des motifs corrélés rares : Application à la détection d'intrusions\\
}

\auteur{Souad Bouasker, Tarek Hamrouni, Sadok Ben Yahia}
\affiliation{Département des Sciences de l'Informatique, Faculté des Sciences de Tunis, Tunisie\\
         \{tarek.hamrouni, sadok.benyahia\}@fst.rnu.tn}
\nomcourt{S. Bouasker, T. Hamrouni et S. Ben Yahia}
\titrecourt{Fouille d'une représentation concise des motifs corrélés rares}

\resume{
La fouille des motifs corrélés qui sont très peu fréquents est une problématique de plus en plus intéressante dans la fouille de données. Dans ce cadre, les motifs corrélés rares selon la mesure de corrélation \textit{bond} ont été étudiés dans un récent travail. La représentation concise exacte $\mathcal{RMCR}$ de l'ensemble de ces motifs a été alors proposée. Toutefois, aucun algorithme n'a été proposé pour extraire cette représentation et aucune évaluation expérimentale de cette représentation n'a été réalisée. Dans ce papier $^\textsc{(}$\footnote{Ce travail propose une version étendue de l'article "Algorithmes d'extraction et d'interrogation d'une représentation concise exacte des motifs corrélés rares : Application à la détection d'intrusions", In Actes de la 12${i\grave{e}me}$ Conférence Internationale Francophone Extraction et Gestion des Connaissances (EGC 2012), 31 Janvier - 03 Février 2012, Bordeaux, France.}$^\textsc{)}$, nous introduisons l'algorithme \textsc{RcprMiner} d'extraction de $\mathcal{RMCR}$. Nous présentons également l'algorithme \textsc{EstMCR}, d'interrogation de cette représentation ainsi que l'algorithme \textsc{RegenerationMCR} de dérivation de tous les motifs corrélés rares à partir de $\mathcal{RMCR}$. L'étude expérimentale réalisée montre des taux de compacité intéressants offerts par cette représentation. En outre, le processus de classification basé sur les règles génériques corrélées rares, dérivées à partir de $\mathcal{RMCR}$, a prouvé l'utilité de l'approche proposée dans le cadre de la détection d'intrusions.}

\summary{
Correlated rare pattern mining is an interesting issue in Data mining. In this respect, the set of correlated rare patterns w.r.t. to the \textit{bond} correlation measure was studied in a recent work, in which the $\mathcal{RCPR}$ concise exact representation of the set of correlated rare patterns was proposed. However, none algorithm was proposed in order to mine this representation and none experiment was carried out to evaluate it. In this paper, we introduce the new \textsc{RcprMiner} algorithm allowing an efficient extraction of $\mathcal{RCPR}$. We also present the \textsc{IsRCP} algorithm allowing the query of the $\mathcal{RCPR}$ representation in addition to the \textsc{RCPRegeneration} algorithm allowing the regeneration of the whole set $\mathcal{RCP}$ of rare correlated patterns starting from this representation. The carried out experiments highlight interesting compactness rates offered by $\mathcal{RCPR}$. The effectiveness of the proposed classification method, based on generic rare correlated association rules derived from $\mathcal{RCPR}$, has also been proved in the context of intrusion detection.}

\begin{document}
\section{Introduction et motivations} \label{section_intro}
L'intégration des mesures de corrélation lors de l'extraction des motifs rares est une piste prometteuse en fouille de données. Elle permet, d'une part, d'améliorer la qualité des connaissances extraites en ayant un ensemble plus réduit contenant des motifs intéressants qui sont rares mais fortement corrélés. D'autre part, ceci renforce la qualité des règles d'association dérivées à partir de ces motifs corrélés rares. Par exemple, le motif composé par les items ``Collier en or'' et ``Boucles d'oreilles'' ou aussi celui composé de ``Télévision'' et ``Lecteur DVD'' correspondent à des motifs fortement corrélés mais peu fréquents dans les transactions d'une grande surface, et peuvent ainsi être omis dans un processus de fouille classique des motifs fréquents. L'utilité de tels motifs a été étudiée dans divers travaux tels que \citep{Kim-pkdd2011,Omie03,borgelt,surana2010,Xiong06hypercliquepattern}.

Dans la littérature, diverses approches d'extraction traitant de cette problématique ont été ainsi proposées. Nous citons, par exemple, l'approche décrite dans \citep{thomo_2010}. Cette dernière est basée sur l'idée naïve d'extraire l'ensemble de tous les motifs fréquents pour un seuil minimal de support conjonctif, \textit{minsupp}, très bas puis de filtrer ces motifs récupérés par la contrainte de corrélation. Cette opération est très coûteuse en temps de traitement et en consommation de la mémoire à cause de l'explosion du nombre de candidats à évaluer. Une autre stratégie d'extraction des motifs rares fortement corrélés, consiste à extraire l'ensemble de tous les motifs corrélés sans aucune intégration de la contrainte de support. Cette idée permet de récupérer les motifs corrélés qui sont très peu fréquents, cependant, elle est aussi coûteuse. Nous citons dans ce cadre les approches proposées dans \citep{ma-icdm2001} et \citep{Cohen_mcr_2000}. Il est important de noter que la contrainte monotone de rareté n'a été jamais incorporée dans la fouille afin de récupérer l'ensemble total des motifs rares fortement corrélés. En effet, les algorithmes proposés dans \citep{Brin97} et \citep{grahne_correlated_2000}, bien qu'ils permettent d'intégrer cette contrainte dans le processus de fouille, se limitent à l'extraction d'un sous-ensemble restreint composé uniquement des motifs minimaux valides \textit{c.-à.-d.} satisfaisant l'ensemble de contraintes posées.

Dans \citep{rnti2012}, la représentation concise $\mathcal{RMCR}$ des motifs corrélés rares associés à la mesure de corrélation \textit{bond} \citep{Omie03} a été proposée. D'un point de vue qualitatif, le choix de cette mesure a été effectué sur la base d'une étude détaillée de la littérature montrant son utilité dans le maintien de motifs intéressants \citep{tsi2012,borgelt,surana2010}. D'un point de vue quantitatif, basée sur la notion clé de classe d'équivalence, cette représentation permet de ne présenter à l'utilisateur qu'un ensemble réduit de motifs tout en offrant la possibilité de dériver, si besoin, ceux non-retenus d'une manière simple et efficace. Toutefois, aucun algorithme n'a été proposé auparavant afin d'extraire une telle représentation. \`A cet égard, nous proposons, dans ce papier, un nouvel algorithme de fouille de la représentation $\mathcal{RMCR}$. Les algorithmes d'interrogation de cette représentation et de dérivation de l'ensemble total des motifs corrélés rares sont aussi présentés. En plus, nous décrivons les résultats obtenus prouvant les taux de compacité importants offerts par $\mathcal{RMCR}$ ainsi que son apport dans la détection d'intrusions. Il est important de noter qu'aucun de ces algorithmes n'a été proposé et aucune expérimentation n'a été réalisée dans \citep{rnti2012}.

Le reste de ce papier est organisé comme suit : la section suivante présente l'ensemble des motifs corrélés rares et la représentation concise $\mathcal{RMCR}$ qui lui est associée. Dans la section \ref{section_algo}, nous introduisons l'algorithme \textsc{RcprMiner} d'extraction de $\mathcal{RMCR}$. La section \ref{section_interrogation} est dédiée à la présentation de l'algorithme d'interrogation de $\mathcal{RMCR}$, tandis que la section \ref{section_regener} décrit le processus de régénération de l'ensemble de tous les motifs corrélés rares à partir de $\mathcal{RMCR}$. L'étude expérimentale est détaillée dans la section \ref{section_XP}. L'application de la représentation $\mathcal{RMCR}$ dans le cadre de la détection d'intrusions est illustrée dans la section \ref{section_IDS}. La conclusion et les perspectives de travaux futurs sont récapitulées dans la section \ref{section_cl}.


\section{Motifs corrélés rares : Définition et représentation concise} \label{section_rappel}
\subsection{Notions de base}

Nous commençons par définir d'abord une base de transactions.
\begin{definition} \label{definitionbasetransactions} \textsc{(}\textbf{Base de transactions}\textsc{)} Une base de transactions est représentée sous la forme d'un triplet $\mathcal{D}$ = \textsc{(}$\mathcal{T}$, $\mathcal{I}$, $\mathcal{R}$\textsc{)} dans lequel $\mathcal{T}$ et $\mathcal{I}$ sont, respectivement, des ensembles finis de transactions \textsc{(}ou objets\textsc{)} et d'items \textsc{(}ou attributs\textsc{)}, et $\mathcal{R}$ $\subseteq$ $\mathcal{T} \times \mathcal{I}$ est une relation binaire entre les transactions et les items. Un couple \textsc{(}$t$, $i$\textsc{)} $\in$ $\mathcal{R}$ dénote le fait que la transaction $t$ $\in$ $\mathcal{T}$ contient l'item $i$ $\in$ $\mathcal{I}$.
\end{definition}
Dans ce travail, nous nous sommes principalement intéressés aux itemsets comme classe de motifs. Nous distinguons trois types de supports correspondants à tout motif non vide $X$ :\\
- \textbf{\textit{Le support conjonctif :}} \textit{SConj}\textsc{(}$X$\textsc{)} = $\mid$$\{$$t$ $\in$ $\mathcal{T}$ $\mid$ $\forall$ $i$ $\in$ $X$ : \textsc{(}$t$, $i$\textsc{)} $\in$ $\mathcal{R}$$\}$$\mid$\\
- \textbf{\textit{Le support disjonctif :}} \textit{SDisj}\textsc{(}$X$\textsc{)} = $\mid$$\{$$t$ $\in$ $\mathcal{T}$ $\mid$ $\exists$ $i$ $\in$ $X$ : \textsc{(}$t$, $i$\textsc{)} $\in$ $\mathcal{R}$$\}$$\mid$\\
- \textbf{\textit{Le support négatif :}} \textit{SNeg}\textsc{(}$X$\textsc{)} = $\mid$$\{$$t$ $\in$ $\mathcal{T}$ $\mid$ $\forall$ $i$ $\in$ $X$ : \textsc{(}$t$, $i$\textsc{)} $\notin$ $\mathcal{R}$$\}$$\mid$


\begin{table}[!t]
\begin{center}
\footnotesize{
\begin{tabular}{|c||c|c|c|c|c|c|}
\hline & \texttt{A}  & \texttt{B}  & \texttt{C}  & \texttt{D} & \texttt{E}  \\
\hline\hline
		1 & $\times$  &          &$\times$    & $\times$&         \\
\hline
		2 &           & $\times$ &$\times$    &         & $\times$ \\
\hline
		3 & $\times$  & $\times$ &$\times$    &         & $\times$ \\
\hline
		4 &           & $\times$ &            &         & $\times$ \\
\hline
		5 & $\times$  & $\times$ &$\times$    &         & $\times$  \\
\hline
\end{tabular}}
\end{center}
\caption{Un exemple d'une base de transactions.}\label{Base_transactions}
\end{table}

\begin{exemple}
Considérons la base de transactions illustrée par la table \ref{Base_transactions}. Nous avons \textit{SConj}\textsc{(}\texttt{AD}\textsc{)} = $\mid$$\{$1$\}$$\mid$ = $1$, \textit{SDisj}\textsc{(}\texttt{AD}\textsc{)} = $\mid$$\{$ 1, 3, 5$\}$$\mid$ = $3$, et, \textit{SNeg}\textsc{(}\texttt{AD}\textsc{)} = $\mid$$\{$2, 4$\}$$\mid$ = $2$ $^{\textsc{(}}$\footnote{Nous employons une forme sans séparateur pour les ensembles d'items : par exemple, \texttt{AD} représente l'ensemble $\{$\texttt{A}, \texttt{D}$\}$.}$^{\textsc{)}}$.
\end{exemple}
La fréquence conjonctive \textsc{(}\textit{resp.} disjonctive et négative\textsc{)} est égale au support conjonctif \textsc{(}\textit{resp.} disjonctif et négatif\textsc{)} divisé par $|$$\mathcal{T}$$|$. Dans la suite, nous allons utiliser les supports d'un motif. Comme nous nous intéressons aux motifs corrélés rares associés à la mesure de corrélation \textit{bond} \citep{Omie03}, la définition suivante présente l'expression de \textit{bond} telle que redéfinie dans \citep{tsi2012}. Cette nouvelle expression permet de faire le lien entre la mesure \textit{bond} et les supports conjonctif et disjonctif, cette mesure étant égale au rapport entre ces deux derniers.
\begin{definition}\label{La mesure bond} \textsc{(}\textbf{Mesure \textit{bond}}\textsc{)} Soit un motif non vide $X$ $\subseteq$ $\mathcal{I}$. La mesure \textit{bond} de $X$ est égale à :
\begin{center}
\textit{bond}\textsc{(}$X$\textsc{)} = $\frac{\displaystyle
\textit{SConj}\textsc{(}X\textsc{)}}{\displaystyle
\textit{SDisj}\textsc{(}X\textsc{)}}$
\end{center}
\end{definition}
Ainsi, connaissant la valeur de \textit{bond} et le support conjonctif d'un motif, il est aisé de dériver son support disjonctif et par conséquent son support négatif. Dans la sous-section suivante, nous présentons l'ensemble $\mathcal{MCR}$ des motifs corrélés rares associés à la mesure \textit{bond}.


\subsection{L'ensemble $\mathcal{MCR}$ des motifs corrélés rares}
Les motifs corrélés rares ont été formellement définis dans \citep{rnti2012} comme suit :
\begin{definition}\label{MCRare}\textsc{(}\textbf{Motifs corrélés rares}\textsc{)} Étant donnés les seuils minimaux de support conjonctif et de corrélation \textit{minsupp} et \textit{minbond}, respectivement, l'ensemble $\mathcal{MCR}$ des motifs corrélés rares est : $\mathcal{MCR}$ = $\{$$X$ $\subseteq$ $\mathcal{I}$$|$ \textit{SConj}\textsc{(}$X$\textsc{)} $<$ \textit{minsupp} et \textit{bond}\textsc{(}$X$\textsc{)} $\geq$ \textit{minbond}$\}$.
\end{definition}
\begin{center}
\begin{figure}[htbp]\centering
 \includegraphics[scale = 0.3]{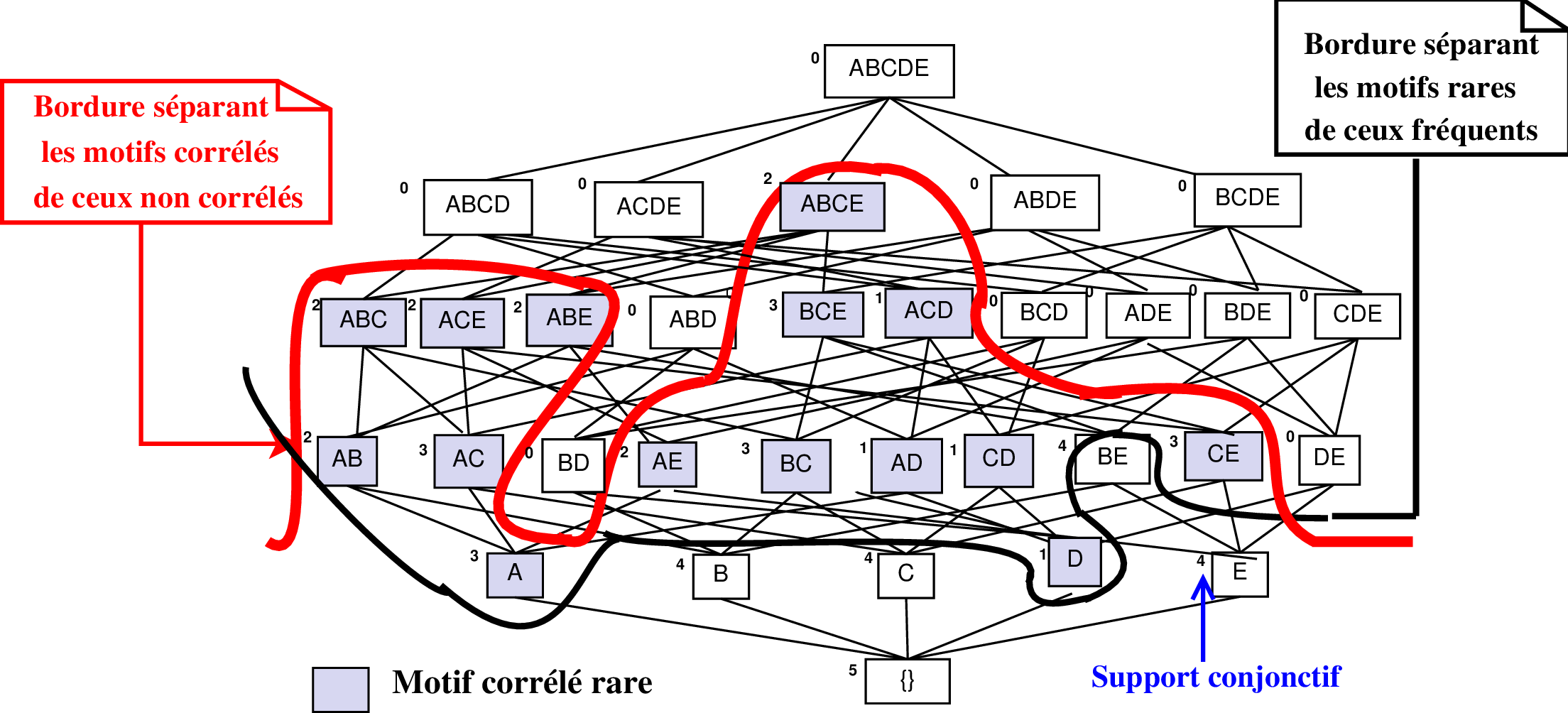}
 \vspace{-0.4cm}
 \caption{Espace des motifs corrélés rares pour \textit{minsupp} = 4 et \textit{minbond} = 0,2.}
 \label{figure_MCR}
\end{figure}
\end{center}

\begin{exemple}\label{exemple_ensemble_MCR}
Considérons la base illustrée par la table \ref{Base_transactions} pour \textit{minsupp} = 4 et \textit{minbond} = 0,2. L'ensemble $\mathcal{MCR}$ est composé des motifs suivants où chaque triplet représente le motif, sa valeur de support conjonctif et sa valeur de \textit{bond} : $\mathcal{MCR}$ = $\{$\textsc{(}\texttt{A}, 3, $\frac{3}{3}$\textsc{)},
\textsc{(}\texttt{D}, 1, $\frac{1}{1}$\textsc{)},
\textsc{(}\texttt{AB}, 2, $\frac{2}{5}$\textsc{)},
\textsc{(}\texttt{AC}, 3, $\frac{3}{4}$\textsc{)},
\textsc{(}\texttt{AD}, 1, $\frac{1}{3}$\textsc{)},
\textsc{(}\texttt{AE}, 2, $\frac{2}{5}$\textsc{)},
\textsc{(}\texttt{BC}, 3, $\frac{3}{5}$\textsc{)},
\textsc{(}\texttt{CD}, 1, $\frac{1}{4}$\textsc{)},
\textsc{(}\texttt{CE}, 3, $\frac{3}{5}$\textsc{)},
\textsc{(}\texttt{ABC}, 2, $\frac{2}{5}$\textsc{)},
\textsc{(}\texttt{ABE}, 2, $\frac{2}{5}$\textsc{)},
\textsc{(}\texttt{ACD}, 1, $\frac{1}{4}$\textsc{)},
\textsc{(}\texttt{ACE}, 2, $\frac{2}{5}$\textsc{)},
\textsc{(}\texttt{BCE}, 3, $\frac{3}{5}$\textsc{)}, \textsc{(}\texttt{ABCE}, 2, $\frac{2}{5}$\textsc{)}$\}$. Comme le montre cette figure, l'ensemble $\mathcal{MCR}$ correspond aux motifs localisés en dessous de la bordure de la contrainte anti-monotone composée des motifs corrélés maximaux, et au dessus de la bordure de la contrainte monotone composée des motifs rares minimaux.
\end{exemple}
L'ensemble $\mathcal{MCR}$ des motifs corrélés rares associés à la mesure \textit{bond} résulte ainsi de la conjonction de deux contraintes de types opposés, à savoir la contrainte anti-monotone de la corrélation et la contrainte monotone de la rareté. Cette nature opposée des contraintes traitées rend complexe la localisation de l'ensemble des motifs corrélés rares. Ceci a motivé les auteurs dans \citep{rnti2012} à introduire la représentation concise exacte $\mathcal{RMCR}$.


\subsection{La représentation concise exacte $\mathcal{RMCR}$}

La représentation concise exacte proposée constitue une réduction sans perte d'informations de l'ensemble $\mathcal{MCR}$. Pour cela, les auteurs ont recouru à la notion de bordure afin de délimiter l'espace associé à l'ensemble $\mathcal{MCR}$ dans le treillis des motifs. Par ailleurs, l'ensemble des motifs corrélés rares a été ainsi partitionné en groupes disjoints, appelés ``classes d'équivalence corrélées rares'' en utilisant l'opérateur de fermeture $f_{bond}$ \citep{tsi2012} associé à la mesure \textit{bond} et défini comme suit.
\begin{definition} \label{fermeture_fbond1} \textsc{(}\textbf{Opérateur $f_{bond}$}\textsc{)} L'opérateur $f_{bond}$ : $\mathcal{P}$\textsc{(}$\mathcal{I}$\textsc{)} $\rightarrow$ $\mathcal{P}$\textsc{(}$\mathcal{I}$\textsc{)} associé à la mesure \textit{bond} est défini comme suit : $f_{bond}$\textsc{(}$X$\textsc{)} =
$X$ $\cup$ $\{$$i$ $\in$ $\mathcal{I}$ $\setminus$ $X$ $|$ \textit{bond}\textsc{(}$X$\textsc{)} = \textit{bond}\textsc{(}$X$ $\cup$ $\{$$i$$\}$\textsc{)}$\}$.
\end{definition}
Chacune des classes d'équivalences induites par l'opérateur $f_{bond}$ regroupe les motifs partageant les mêmes supports conjonctifs, disjonctifs et la même valeur de la mesure de corrélation \textit{bond}. Les éléments maximaux des classes d'équivalence corrélées rares composent l'ensemble $\mathcal{MFCR}$ des motifs fermés corrélés rares et les éléments minimaux composent l'ensemble $\mathcal{MMCR}$ des motifs minimaux corrélés rares, qui ont été définis comme suit.
\vspace{-0.1cm}
\begin{definition}\label{MFCR} \textsc{(}\textbf{Motifs fermés corrélés rares}\textsc{)} L'ensemble $\mathcal{MFCR}$ des motifs fermés corrélés rares est défini par : $\mathcal{MFCR}$ = $\{$$X$ $\in$ $\mathcal{MCR}$$|$ $\forall$ $X_{1}$ $\supset$ $X$ : \textit{bond}\textsc{(}$X$\textsc{)} $>$ \textit{bond}\textsc{(}$X_{1}\textsc{)}\}$
\end{definition}
\begin{definition}\label{MMCR} \textsc{(}\textbf{Motifs minimaux corrélés rares}\textsc{)} L'ensemble $\mathcal{MMCR}$ des motifs minimaux corrélés rares est défini par : $\mathcal{MMCR}$ = $\{$$X$ $\in$ $\mathcal{MCR}$$|$ $\forall$ $X_{1}$ $\subset$ $X$ : \textit{bond}\textsc{(}$X$\textsc{)} $<$ \textit{bond}\textsc{(}$X_{1}\textsc{)}\}$.
\end{definition}
\begin{exemple}\label{exemple_MMCR_et_MFCR}
Soit la base illustrée par la table \ref{Base_transactions} pour \textit{minsupp} = 4 et \textit{minbond} = 0,2.
Nous avons, par exemple, $f_{bond}$\textsc{(}\texttt{AB}\textsc{)} = \texttt{ABCE}, l'ensemble $\mathcal{MFCR}$ = $\{$\texttt{A},  \texttt{D}, \texttt{AC}, \texttt{AD}, \texttt{ACD}, \texttt{BCE}, \texttt{ABCE}$\}$.
Par ailleurs, l'ensemble $\mathcal{MMCR}$ = $\{$\texttt{A}, \texttt{D}, \texttt{AB}, \texttt{AC}, \texttt{AD}, \texttt{AE}, \texttt{BC}, \texttt{CD}, \texttt{CE}$\}$. Il est intéressant de remarquer que les motifs \texttt{A}, \texttt{D}, \texttt{AC} et \texttt{AD} sont à la fois fermés et minimaux.
\end{exemple}
En se basant sur ces deux ensembles précédents, la représentation $\mathcal{RMCR}$ de l'ensemble $\mathcal{MCR}$ a été proposée.
\begin{definition}\label{rmcr} \textsc{(}\textbf{Représentation $\mathcal{RMCR}$}\textsc{)} 
La représentation $\mathcal{RMCR}$ est définie comme suit : $\mathcal{RMCR}$ = $\mathcal{MFCR}$ $\cup$ $\mathcal{MMCR}$.
\end{definition}
\begin{exemple} \label{ExpRMCR}
Considérons la base de transactions donnée dans la table \ref{Base_transactions}, pour \textit{minsupp} = 4 et \textit{minbond} = 0,2. La représentation $\mathcal{RMCR}$ = $\{$\textsc{(}\texttt{A}, $3$, $\frac{3}{3}$\textsc{)},
\textsc{(}\texttt{D}, $1$, $\frac{1}{1}$\textsc{)},
\textsc{(}\texttt{AB}, $2$, $\frac{2}{5}$\textsc{)},
\textsc{(}\texttt{AC}, $3$, $\frac{3}{4}$\textsc{)},
\textsc{(}\texttt{AD}, $1$, $\frac{1}{3}$\textsc{)},
\textsc{(}\texttt{AE}, $2$, $\frac{2}{5}$\textsc{)},
\textsc{(}\texttt{BC}, $3$, $\frac{3}{5}$\textsc{)},
\textsc{(}\texttt{CD}, $1$, $\frac{1}{4}$\textsc{)},
\textsc{(}\texttt{CE}, $3$, $\frac{3}{5}$\textsc{)},
\textsc{(}\texttt{ACD}, $1$, $\frac{1}{4}$\textsc{)},
\textsc{(}\texttt{BCE}, $3$, $\frac{3}{5}$\textsc{)},
\textsc{(}\texttt{ABCE}, $2$, $\frac{2}{5}$\textsc{)}$\}$.
\end{exemple}
Cette représentation a été prouvée dans \citep{rnti2012} comme étant exacte, \textit{c.-à.-d.} permettant la régénération de tous les motifs corrélés rares sans perte d'informations. Par ailleurs, sa taille ne dépasse jamais celle de $\mathcal{MCR}$. En effet, $\mathcal{RMCR}$ = $\mathcal{MFCR}$ $\cup$ $\mathcal{MMCR}$ $\subseteq$ $\mathcal{MCR}$.

Nous introduisons, dans ce qui suit, l'algorithme \textsc{RcprMiner} permettant l'extraction de la représentation
concise exacte $\mathcal{RMCR}$.
\section{Algorithme \textsc{RcprMiner} d'extraction de $\mathcal{RMCR}$} \label{section_algo}

\subsection{Description et pseudo code de l'algorithme \textsc{RcprMiner}}

L'algorithme \textsc{RcprMiner}$^\textsc{(}$\footnote{Acronyme de Rare Correlated Patterns Representation Miner.}$^\textsc{)}$, dont le pseudo-code est donné par l'algorithme \ref{AlgoRCPR}, prend en entrée une base de transactions $\mathcal{D}$, un seuil minimal de support conjonctif \textit{minsupp} ainsi qu'un seuil
minimal de corrélation \textit{minbond}. Cet algorithme permet de déterminer, à partir du contexte $\mathcal{D}$, la représentation $\mathcal{RMCR}$ composée de l'ensemble $\mathcal{MMCR}$ des motifs minimaux corrélés rares  et de l'ensemble $\mathcal{MFCR}$ des motifs fermés corrélés rares munis de leurs supports conjonctifs et de leurs valeurs de la mesure \textit{bond}.

Le déroulement de l'algorithme \textsc{RcprMiner} est illustrée par la figure 2. Cet algorithme se réalise en deux principales étapes. La première étape est dédiée à l'extraction, à partir de $\mathcal{D}$, de l'ensemble $\mathcal{MCM}ax$ des motifs corrélés maximaux grâce à la procédure \textsc{Extraction\_MCMax} \textsc{(}\textit{cf.} ligne 4\textsc{)}. Cette étape consiste à résoudre un problème classique permettant le repérage des éléments maximaux d'une théorie, les motifs maximaux associés à l'ensemble des motifs corrélés dans notre cas.
\begin{center}
\begin{figure}[htbp]\centering
 \includegraphics[scale = 0.25]{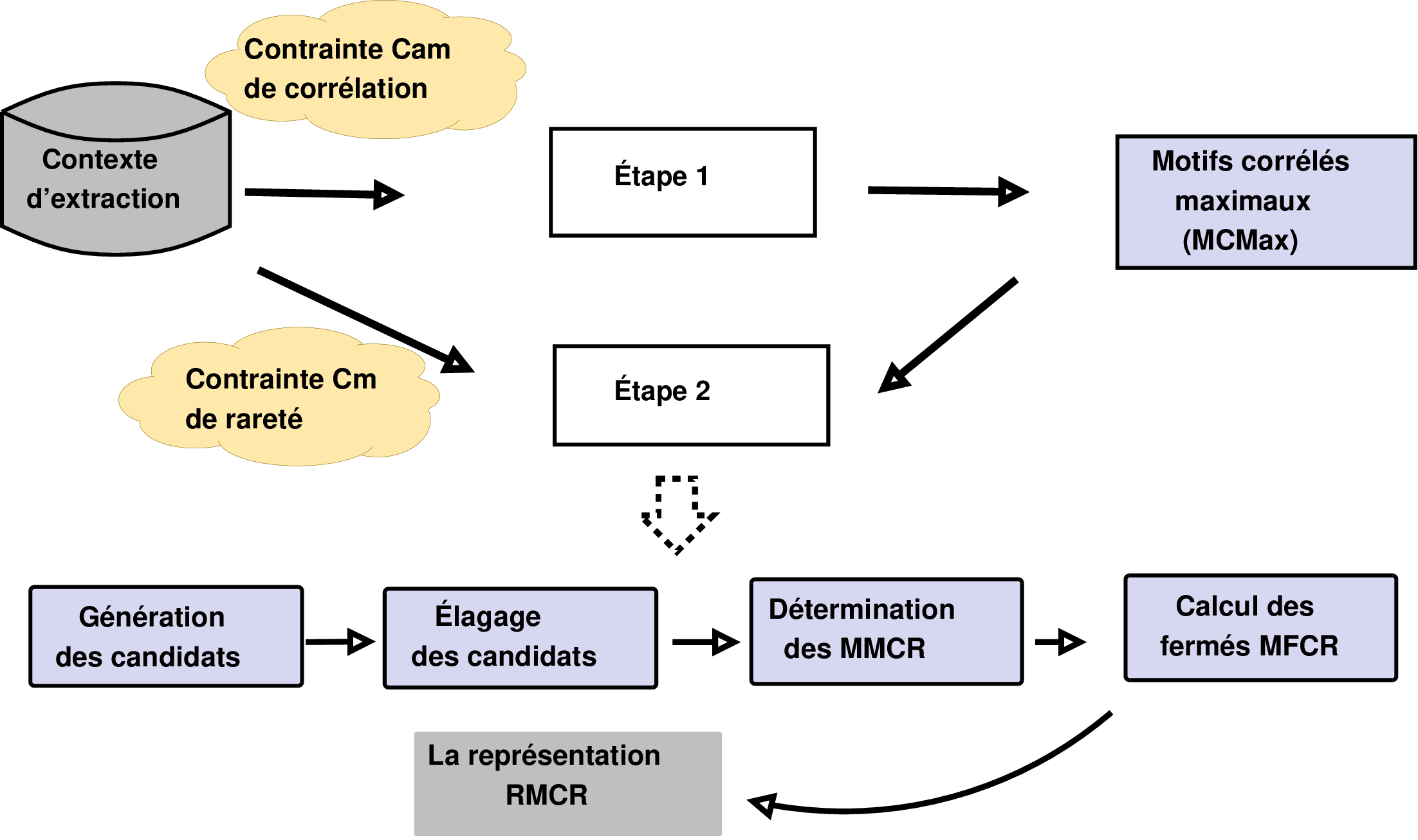}
 \vspace{-0.4cm}
 \caption{Schéma illustratif de déroulement de l'algorithme \textsc{RcprMiner}.}
 \label{figure_Algo}
\end{figure}
\end{center}
La deuxième étape consiste à intégrer la contrainte de rareté ainsi que les motifs corrélés maximaux précédemment extraits dans un processus de fouille de la représentation $\mathcal{RMCR}$. \`A chaque itération de cette deuxième étape, l'ensemble $\mathcal{C}$$and\mathcal{P}$$_{n}$ est composé des candidats potentiels de taille $n$ générés, moyennant la procédure \textsc{Apriori\_Gen} \textsc{(}\textit{cf.} ligne 15\textsc{)}, à partir des candidats retenus de taille ($n-1$). Les candidats de l'ensemble $\mathcal{C}$$and\mathcal{P}$$_{n}$ seront ainsi élagués \textsc{(}\textit{cf.} ligne 11\textsc{)} selon différentes stratégies d'élagage. Les éléments maintenus seront insérés dans l'ensemble $\mathcal{C}$$and_{n}$. Les stratégies d'élagage adoptées correspondent à :

\textsc{(}\textbf{\textit{i}}\textsc{)} \textbf{L'élagage de tout candidat inclus dans un motif corrélé maximal fréquent}, puisqu'il sera corrélé fréquent d'après la propriété de l'idéal d'ordre des motifs corrélés fréquents \textsc{(}la contrainte de corrélation étant anti-monotone\textsc{)}.

\textsc{(}\textbf{\textit{ii}}\textsc{)} \textbf{L'élagage de tout candidat non inclus dans un motif corrélé maximal rare}, puisqu'il ne sera pas corrélé.

\textsc{(}\textbf{\textit{iii}}\textsc{)} \textbf{L'élagage par rapport à la propriété d'idéal d'ordre des motifs minimaux corrélés} : en effet, les motifs minimaux corrélés vérifient la propriété de l'idéal d'ordre. Ainsi, tout candidat minimal corrélé possédant un sous-ensemble non minimal corrélé, sera élagué vu qu'il ne sera pas un motif minimal corrélé.


Notons que tout candidat potentiel inclus dans un motif corrélé maximal rare est forcément corrélé. Toutefois, nous ne pouvons rien confirmer quant à sa rareté. \`A cet égard, il sera retenu dans l'ensemble $\mathcal{C}$$and_{n}$ et son statut de fréquence sera vérifié grâce à la procédure \textsc{Extraction\_MMCR\_MFCR}, \textsc{(}\textit{cf.} ligne 13\textsc{)}, dont le pseudo code est donné par l'algorithme \ref{calculSuppF}. Cette procédure permet de déterminer les motifs minimaux corrélés rares à partir des candidats retenus dans l'ensemble $\mathcal{C}$$and_{n}$. Pour cela, la valeur de \textit{bond} de chaque candidat sera comparée à celles de ses sous-ensembles directs pour déterminer s'il est minimal dans sa classe d'équivalence ou non. En effet, tout candidat ayant la même valeur de corrélation qu'un de ses sous-ensembles n'est pas minimal de sa classe. Les motifs minimaux corrélés rares identifiés seront ainsi insérés dans l'ensemble $\mathcal{MMCR}$. Une fois les minimaux repérés, leurs fermetures sont calculées et insérées dans l'ensemble $\mathcal{MFCR}$. Par ailleurs, dans l'ensemble $\mathcal{C}$$and_{n}$, seuls les candidats minimaux de leurs classes d'équivalence seront maintenus. Ceci permet d'utiliser $\mathcal{C}$$and_{n}$ dans l'élagage des candidats potentiels de taille \textsc{(}$n$ + 1\textsc{)} \textsc{(}\textit{cf.} la stratégie d'élagage \textsc{(}\textit{iii}\textsc{)} de la ligne 11\textsc{)}.

\begin{exemple} Considérons la base de transactions donnée par la table \ref{Base_transactions}. L'algorithme \textsc{RcprMiner} se déroule de la manière suivante pour \textit{minsupp} = 3 et \textit{minbond} = 0,20.
Nous avons, initialement, l'ensemble $\mathcal{MCM}ax$ = $\{$\textsc{(}\texttt{ACD}, 1, $\frac{1}{4}$\textsc{)}, \textsc{(}\texttt{ABCE}, 2, $\frac{2}{5}$\textsc{)}$\}$.
\'Etant donné que tous les motifs de cet ensemble sont rares, nous avons donc $\mathcal{MCM}ax$$\mathcal{R}$ = $\{$\textsc{(}\texttt{ACD}, 1, $\frac{1}{4}$\textsc{)}, \textsc{(}\texttt{ABCE}, 2, $\frac{2}{5}$\textsc{)}$\}$. Ensuite, nous avons $\mathcal{C}$$and$$\mathcal{P}$$_{1}$ =  $\{$\texttt{A}, \texttt{B}, \texttt{C}, \texttt{D}, \texttt{E}$\}$. Il en dérive, $\mathcal{MMCR}$$_{1}$ = $\{$\textsc{(}\texttt{D}, 1, $\frac{1}{1}$\textsc{)}$\}$ et $\mathcal{MFCR}$$_{1}$ = $\{$\textsc{(}\texttt{D}, 1, $\frac{1}{1}$\textsc{)}$\}$. L'ensemble $\mathcal{C}$$and$$\mathcal{P}$$_{2}$ est ensuite généré : $\mathcal{C}$$and$$\mathcal{P}$$_{2}$ =  $\{$\texttt{AB}, \texttt{AC}, \texttt{AD}, \texttt{AE}, \texttt{BC}, \texttt{BE}, \texttt{BD}, \texttt{CE}, \texttt{CD}, \texttt{DE}$\}$. Suite à l'application des stratégies d'élagage, nous avons, $\mathcal{MMCR}$$_{2}$ = $\{$\textsc{(}\texttt{AB}, 2, $\frac{2}{5}$\textsc{)}, \textsc{(}\texttt{AE}, 2, $\frac{2}{5}$\textsc{)}, \textsc{(}\texttt{AD}, 1, $\frac{1}{3}$\textsc{)},  \textsc{(}\texttt{CD}, 1, $\frac{1}{4}$\textsc{)}$\}$.
Les motifs fermés associés à ces minimaux, à savoir \textsc{(}\texttt{AD}, 1, $\frac{1}{3}$\textsc{)}, \textsc{(}\texttt{ACD}, 1, $\frac{1}{4}$\textsc{)} et \textsc{(}\texttt{ABCE}, 2, $\frac{2}{5}$\textsc{)}, sont alors ajoutés à $\mathcal{MFCR}$. Dans la troisième itération, nous avons $\mathcal{C}$$and$$\mathcal{P}$$_{3}$ = $\{$\texttt{ABC}, \texttt{ABD}, \texttt{ABE}, \texttt{ACD}, \texttt{ACE}, \texttt{ADE}, \texttt{BCD}, \texttt{BCE}, \texttt{CDE}$\}$. Aucun de ces candidats n'est minimal rare corrélé. Ainsi, $\mathcal{MMCR}$$_{3}$ = $\{$$\emptyset$$\}$. L'ensemble des candidats $\mathcal{C}$$and$$\mathcal{P}$$_{4}$ est par conséquent vide. Ainsi, les itérations prennent fin donnant ainsi comme résultat les motifs minimaux corrélés rares $\mathcal{MMCR}$ = $\{$\textsc{(}\texttt{D}, 1, $\frac{1}{1}$\textsc{)},
\textsc{(}\texttt{AB}, 2, $\frac{2}{5}$\textsc{)},
\textsc{(}\texttt{AE}, 2, $\frac{2}{5}$\textsc{)},
\textsc{(}\texttt{AD}, 1, $\frac{1}{3}$\textsc{)},
\textsc{(}\texttt{CD}, 1, $\frac{1}{4}$\textsc{)}$\}$
et leurs fermés  $\mathcal{MFCR}$ = $\{$\textsc{(}\texttt{D}, 1, $\frac{1}{1}$\textsc{)},
\textsc{(}\texttt{AD}, 1, $\frac{1}{3}$\textsc{)},
\textsc{(}\texttt{ACD}, 1, $\frac{1}{4}$\textsc{)},
\textsc{(}\texttt{ABCE}, 2, $\frac{2}{5}$\textsc{)}$\}$.
\end{exemple}


\incmargin{0.5em} \linesnumbered
\begin{algorithm}[htbp]\label{AlgoRCPR}
 \small{ \small{ \caption{\textsc{RcprMiner}}}
   \SetVline
	\setnlskip{-3pt}
\Donnees{Une base de transactions $\mathcal{D}$ = $\textsc{(}$$\mathcal{T},\mathcal{I},\mathcal{R}$$\textsc{)}$, \textit{minbond}, et \textit{minsupp}.}
 \Res{La représentation concise exacte $\mathcal{RMCR}$ = $\mathcal{MMCR}$ $\cup$ $\mathcal{MFCR}$.}
 \Deb{
$\mathcal{RMCR}$ := $\emptyset$; $\mathcal{C}$$and_{0}$ := \{$\emptyset$\}\;

/* \textbf{La première étape} */\\

$\mathcal{MCM}$$ax$ := \textsc{Extraction\_MCMax}\textsc{(}$\mathcal{D}$,  \textit{minbond}\textsc{)}\;
/* \textbf{La deuxième étape} */\\

$\mathcal{MCM}$$ax$$\mathcal{F}$ := $\{$$X$ $\in$  $\mathcal{MCM}$$ax$ $\mid$  $X$.\textit{SConj} $\geq$ \textit{minsupp}$\}$ /* $X$.\textit{SConj} correspond au support conjonctif de $X$ */\;

$\mathcal{MCM}$$ax$$\mathcal{R}$ := $\{$$X$ $\in$  $\mathcal{MCM}$$ax$ $\mid$  $X$.\textit{SConj} $<$ \textit{minsupp}$\}$ \label{getmcmR}\;

$\mathcal{C}$$and$$\mathcal{P}$$_{1}$ := $\{$$i$ $|$ $i$ $\in$ $\mathcal{I}$$\}$ \label{setcand1} /* $\mathcal{C}$$and$$\mathcal{P}$$_{n}$ correspond aux candidats potentiels de taille $n$ */\;

\Tq{\textsc{(}$\mathcal{C}$$and$$\mathcal{P}$$_{n}$ $\neq$
$\emptyset$\textsc{)} \label{iter1}}
{
/* \textit{\'Elagage des candidats potentiels} */ \\

$\mathcal{C}$$and_{n}$ := $\mathcal{C}$$and$$\mathcal{P}$$_{n}$ $\backslash$ $\{$$X_{n}$ $\in$ $\mathcal{C}$$and$$\mathcal{P}$$_{n}$
$|$ \textsc{(}$\exists$ $Z$ $\in$ $\mathcal{MCM}$ax$\mathcal{F}$ : $X_{n}$ $\subseteq$ $Z$\textsc{)} ou \textsc{(}$\nexists$ $Z$ $\in$ $\mathcal{MCM}$ax$\mathcal{R}$ : $X_{n}$ $\subseteq$ $Z$\textsc{)} ou \textsc{(}$\exists$ $Y_{n-1}$ $\subset$ $X_{n}$ : $Y_{n-1}$ $\notin$
$\mathcal{C}$$and$$_{n-1}$\textsc{)}$\}$\label{prune3}\;

/* \textit{Détermination des motifs minimaux corrélés rares de taille $n$ et calcul de leurs fermetures} */ \\

$\mathcal{RMCR}$ := $\mathcal{RMCR}$ $\cup$ \textsc{Extraction\_MMCR\_MFCR}\textsc{(}$\mathcal{D}$, $\mathcal{C}$$and_{n}$, \textit{minsupp}\textsc{)}\label{appelfct2}\;

$n$ := $n$ +$1$\;

$\mathcal{C}$$and$$\mathcal{P}$$_{n}$ :=
\textsc{Apriori\_Gen}\textsc{(}$\mathcal{C}$$and_{n-1}$\textsc{)} \label{setcand3}\;
}
\Retour{$\mathcal{RMCR}$\;} \label{output1}}}
\end{algorithm}
\decmargin{0.5em}
\incmargin{0.5em}
\linesnumbered
\begin{algorithm}[!t]\label{calculSuppF}
 \small{\small{ \caption{\textsc{Extraction\_MMCR\_MFCR}}}
   \SetVline
	\setnlskip{-3pt}
\Donnees{La base de transactions $\mathcal{D}$, l'ensemble $\mathcal{C}$$and_{n}$ des motifs candidats de taille $n$, et le seuil minimal de support \textit{minsupp}.} \Res{L'ensemble $\mathcal{MMCR}$$_{n}$ des motifs minimaux corrélés rares de taille $n$ et l'ensemble $\mathcal{MFCR}$ des motifs fermés corrélés rares.
L'ensemble $\mathcal{C}$$and_{n}$ contenant uniquement les motifs minimaux corrélés.} \Deb{
\PourCh{\textsc{(}Transaction $T$ de $\mathcal{D}$\textsc{)}}
	 {
	   \PourCh{\textsc{(}$X_{n}$ $\in$ $\mathcal{C}$$and_{n}$\textsc{)}}
	   {
          $\omega$ := $X_{n}$ $\cap$ $X$  /* $X$ corresponds aux items constituant la transaction $T$ */\;
           \Si {\textsc{(}$\omega$ $=$ $\emptyset$\textsc{)}}
             {
                $X_{n}$.\textit{$CmpDisj$} :=  $X_{n}$.\textit{$CmpDisj$} $\cup$ $X$ /* $X_{n}$.\textit{$CmpDisj$} englobe les items qui apparaissent dans les transactions ne contenant aucun item de $X_{n}$, par conséquent, ces items n'appartiennent donc pas à la fermeture disjonctive du candidat $X_{n}$.*/ \label{majcmpdisj}\;

             }             { $X_{n}$.\textit{SDisj} :=  $X_{n}$.\textit{SDisj} + 1\label{majdisj} /* $X_{n}$.\textit{SDisj} correspond au support disjonctif de $X_{n}$ */\;
                 \Si {\textsc{(}$\omega$ $=$ $X_{n}$\textsc{)}\label{si-egal}}
                 {
                    $X_{n}$.\textit{SConj} :=  $X_{n}$.\textit{SConj} + 1\label{majconj} \;
                     \Si {$X_{n}$.$f_{c}$ = $\emptyset$}
  	                    {$X_{n}$.$f_{c}$ := $\omega$\label{majfc1}\;
  	                    }
  	                  {$X_{n}$.$f_{c}$ := $X_{n}$.$f_{c}$ $\cap$ $\omega$\label{majfc2}\;}
                 }
             }
        }
	  }	
	\PourCh{\textsc{(}$X_{n}$ $\in$ $\mathcal{C}$$and_{n}$\textsc{)}}
	   {
  $X_{n}$.\textit{bond} := $\frac{\displaystyle I.\textit{SConj}}
 {\displaystyle I.\textit{SDisj}}$ \label{calculbond} /* $X_{n}$ est forcément corrélé puisqu'il est inclus dans un motif corrélé maximal*/ \;
   \Si{\textsc{(}$\exists$ $Y_{n-1}$ $\subset$ $X_{n}$ $|$
  \textit{bond}\textsc{(}$Y_{n-1}$\textsc{)} $=$ \textit{bond}\textsc{(}$X_{n}$\textsc{)}\textsc{)}\label{verifMMC}}
  {$\mathcal{C}$$and_{n}$ := $\mathcal{C}$$and_{n}$ $\backslash$ \{$X_{n}$\} /* $X_{n}$ n'est pas un motif minimal corrélé, il est donc élagué de l'ensemble  $\mathcal{C}$$and_{n}$ et ne sera plus utilisé pour la génération de nouveaux candidats */\;}
{ 	  \Si{\textsc{(}$X_{n}$.\textit{SConj} $<$
 \textit{minsupp}\textsc{)}\label{verifsupp}}
 {/* $X_{n}$ est un motif \textit{minimal corrélé rare} */ \\
    $\mathcal{MMCR}$$_{n}$ := $\mathcal{MMCR}$$_{n}$ $\cup$ \textsc{(}$X_{n}$, $X_{n}$.\textit{SConj}, $X_{n}$.\textit{bond}\textsc{)}\label{majMMCR}\;

    $X_{n}$.$f_{d}$ := $\mathcal{I}$$\setminus$$X_{n}$.$CmpDisj$\label{calculfd}\;      	
  	$X_{n}$.$f_{bond}$ :=  $X_{n}$.$f_{d}$ $\cap$ $X_{n}$.$f_{c}$\label{calculfbond}\;
  	$l$ := $|$$X_{n}$.$f_{bond}$$|$ \label{gettaille}\;
  	$\mathcal{MFCR}$$_{l}$ := $\mathcal{MFCR}$$_{l}$ $\cup$ \textsc{(}$X_{n}.f_{bond}$, $X_{n}$.\textit{SConj}, $X_{n}$.\textit{bond}\textsc{)}\label{majMFCR1}\;
  	
   $\mathcal{MFCR}$ := $\mathcal{MFCR}$ $\cup$ $\mathcal{MFCR}$$_{l}$\label{majMFCR2}\;
  	      }
  }
	}
\Retour{\textsc{(}$\mathcal{MMCR}$$_{n}$ $\cup$ $\mathcal{MFCR}$\textsc{)}\;}\label{output2}}}
\end{algorithm}
\decmargin{0.5em}
\subsection{Preuves théoriques}	
Nous démontrons, dans ce qui suit, les propriétés théoriques de validité et de terminaison de l'algorithme \textsc{RcprMiner}.

\begin{proposition}
L'algorithme \textsc{RcprMiner} génère tous les motifs minimaux et fermés corrélés rares munis de leurs supports conjonctifs et de leurs valeurs de la mesure \textit{bond}.
\end{proposition}

\begin{preuve}
L'algorithme \textsc{RcprMiner} est un algorithme par niveau permettant d'extraire avec exactitude tous les éléments de la représentation $\mathcal{RMCR}$. En effet, lors de la première étape, les motifs corrélés maximaux sont identifiés puis ils sont répartis suivant leur statut de fréquence en des motifs corrélés maximaux fréquents et des motifs corrélés maximaux rares. Ces ensembles de motifs seront utilisés pour l'élagage des candidats. Ensuite, les motifs minimaux corrélés rares de l'ensemble $\mathcal{MMCR}$ seront extraits et leurs fermés respectifs seront calculés et insérés dans l'ensemble $\mathcal{MFCR}$ d'une manière itérative.

En effet, lors de chaque itération, un ensemble de candidats de taille $n$ est généré à partir des candidats de taille $n-1$. Chaque motif candidat doit être inclus dans un motif corrélé maximal rare et ne doit posséder aucun sous-ensemble
non minimal corrélé. Ensuite, les supports conjonctifs, disjonctifs, les fermetures conjonctives et les fermetures disjonctives de tous les candidats seront calculés moyennant un balayage du contexte d'extraction. La valeur de la mesure \textit{bond} est ensuite calculée pour tous les candidats retenus. Par la suite, tout candidat possédant un sous-ensemble de même mesure \textit{bond} que lui sera élagué, vu qu'il n'est pas minimal corrélé.

L'ensemble des candidats englobe, à ce niveau, tous les motifs minimaux corrélés. Ainsi, chaque candidat rare sera inséré dans l'ensemble $\mathcal{MMCR}$$_{n}$ des motifs minimaux corrélés rares de taille $n$.
Par conséquent, le motif fermé par $f_{bond}$ correspondant au motif minimal corrélé rare en cours, sera calculé. Il résulte, en effet, de l'intersection entre son fermé conjonctif et son fermé disjonctif.
\'Etant donné que les supports conjonctifs, disjonctifs et la mesure \textit{bond} d'un fermé sont égaux à ceux du motif minimal correspondant, alors nous déduisons que les caractéristiques de chaque fermé par l'opérateur de fermeture $f_{bond}$ sont attribués d'une manière exacte.  Ainsi, l'ensemble $\mathcal{MMCR}_{n}$ ne contient que les motifs minimaux corrélés rares de taille $n$ et l'ensemble $\mathcal{MFCR}_{n}$ ne contient que les fermés corrélés rares de taille $n$.

L'algorithme marque sa fin d'exécution lorsqu'il n'y a plus de motifs candidats à générer. \`A la fin de cette étape l'ensemble $\mathcal{MMCR}$ est composé de tous les motifs qui sont minimaux corrélés rares et leurs fermés respectifs sont inclus dans l'ensemble $\mathcal{MFCR}$.

Nous concluons que l'algorithme \textsc{RcprMiner} permet d'extraire avec exactitude tous les éléments des ensembles $\mathcal{MMCR}$ et $\mathcal{MFCR}$ munis de leurs supports conjonctifs et de leurs valeurs de la mesure \textit{bond}.
Cet algorithme est donc valide et complet.
\end{preuve}

\begin{proposition}
L'algorithme \textsc{RcprMiner} se termine correctement.
\end{proposition}
\begin{preuve}
Le nombre des motifs générés par \textsc{RcprMiner} est fini. En effet, le nombre de motifs candidats pouvant être générés à partir d'un contexte d'extraction ayant $n$ items distincts, est égal au plus à $2^{n}$. De plus, le nombre d'opérations effectuées, afin de traiter chaque candidat est fini. Par conséquent, l'algorithme \textsc{RcprMiner} se termine correctement.
\end{preuve}

Ainsi, nous avons démontré les propriétés théoriques de validité et de terminaison de l'algorithme \textsc{RcprMiner} d'extraction de la représentation $\mathcal{RMCR}$. Dans la section suivante, nous introduisons l'algorithme \textsc{EstMCR} d'interrogation de cette représentation.
\section{Algorithme \textsc{EstMCR} d'interrogation de $\mathcal{RMCR}$} \label{section_interrogation}

L'interrogation de la représentation permet de déterminer pour un motif donné s'il est corrélé rare. Si c'est le cas, alors les valeurs de son support conjonctif, disjonctif, négatif, ainsi que la valeur de sa mesure \textit{bond}, seront régénérées grâce à la représentation $\mathcal{RMCR}$. Ceci est réalisé moyennant l'algorithme \textsc{EstMCR} dont le pseudo-code est donné par l'algorithme \ref{Rege1}.


L'algorithme \textsc{EstMCR} distingue trois différents cas. Le premier se réalise lorsque le motif considéré appartient à la représentation $\mathcal{RMCR}$. Son support disjonctif et son support négatif seront ainsi aisément dérivés \textsc{(}\textit{cf.} lignes 3--4\textsc{)}. Le deuxième cas se présente lorsque le motif $X$ n'appartient pas à la représentation $\mathcal{RMCR}$ mais il est compris entre deux éléments de cette représentation \textsc{(}\textit{cf.} ligne 7\textsc{)}. Ainsi, le motif fermé associé au motif $X$ correspond au plus petit sur-ensemble, selon l'inclusion ensembliste, appartenant à la représentation $\mathcal{RMCR}$ \textsc{(}\textit{cf.} ligne 8\textsc{)}. Le motif $X$ partage ainsi les mêmes valeurs des différents supports et de \textit{bond} que son fermé \textsc{(}\textit{cf.} ligne 9--12\textsc{)}. Dans le troisième et dernier cas, le motif $X$ n'appartient pas à $\mathcal{RMCR}$ et n'est pas compris entre deux éléments de $\mathcal{RMCR}$. Ce motif n'est en conséquent pas corrélé rare et l'algorithme retourne un résultat vide \textsc{(}\textit{cf.} ligne 15\textsc{)}. Nous illustrons dans la suite par un exemple l'exécution de l'algorithme \textsc{EstMCR}.
\begin{exemple}\label{exp_regeneration1} Soit la représentation $\mathcal{RMCR}$ donnée par l'exemple \ref{ExpRMCR} \textsc{(}\textit{cf.} page \pageref{ExpRMCR}\textsc{)}. Considérons le motif \texttt{ACE}. Nous avons \texttt{AE} $\subseteq$ \texttt{ACE} et \texttt{ACE} $\subseteq$ \texttt{ABCE}. Ainsi, le motif \texttt{ACE} est corrélé rare. Par ailleurs, sa fermeture est \texttt{ABCE}. Par conséquent, \texttt{ACE}.\textit{SConj} = \texttt{ABCE}.\textit{SConj} = 2, \texttt{ACE}.\textit{SDisj} = \texttt{ABCE}.\textit{SDisj} = 5, \texttt{ACE}.\textit{SNeg} = $|\mathcal{T}|$ - \texttt{ACE}.\textit{SDisj} = 5 - 5 = 0 et \texttt{ACE}.\textit{bond} = \texttt{ABCE}.\textit{bond} = $\frac{2}{5}$. Considérons le motif \texttt{BC}, ce dernier n'appartient pas à $\mathcal{RMCR}$ et il n'est pas compris entre deux éléments de la représentation. Ainsi, l'algorithme \textsc{EstMCR} retourne un résultat vide pour indiquer que le motif \texttt{BC} n'est pas un motif corrélé rare.
\end{exemple}


\incmargin{0.5em} \linesnumbered
\begin{algorithm}[h]
 \small{ \small{ \caption{\textsc{EstMCR} \label{Rege1}}}
	\SetVline
	\setnlskip{-3pt}
\Donnees{La représentation $\mathcal{RMCR}$ = $\mathcal{MMCR}$ $\cup$ $\mathcal{MFCR}$, un motif $X$, et le nombre de transactions de la base, \textit{c.-à.-d.}, $\mid$$\mathcal{T}$$\mid$.}
	\Res{Le support conjonctif, disjonctif, négatif et la valeur de la mesure \textit{bond} si le motif $X$ est corrélé rare. Sinon, un résultat vide est retourné.}
\Deb{
 \eSi {\textsc{(}$X$ $\in$ $\mathcal{RMCR}$\textsc{)} \label{Apparrmcr}}
{
 $X$.\textit{SDisj} = $\frac{\displaystyle X.\textit{SConj}}{\displaystyle X.\textit{bond}}$ /* $X$.\textit{SDisj} correspond au support disjonctif de $X$ */ \label{calculdisj} \;
$X$.\textit{SNeg} = $\mid$$\mathcal{T}$$\mid$ $-$ $X$.\textit{SDisj} /* $X$.\textit{SNeg} correspond au support négatif de $X$ */\label{calculneg}\;

   \Retour{ $\{$$X$, $X$.\textit{SConj},
   $X$.\textit{SDisj},
   $X$.\textit{SNeg},
   $X$.\textit{bond}$\}$ \label{res1}}\;
  }
  {
   \eSi{\textsc{(}$\exists$ $Y$, $Z$ $\in$ $\mathcal{RMCR}$ $\mid$ $Y$ $\subset$ $X$ et $X$ $\subset$ $Z$\textsc{)}\label{si1}}
     {
        	
$F$ := $\min_{\subseteq}$$\{$$X_{1}$ $\in$  $\mathcal{RMCR}$ $\mid$  $X$  $\subset$ $X_{1}$$\}$ /* $F$ dénote la fermeture de $X$, repérée étant le plus petit motif par inclusion ensembliste de la représentation englobant $X$ */ \label{fermé}\;

 $X$.\textit{SConj} = $F$.\textit{SConj}\;
 $X$.\textit{bond} = $F$.\textit{bond}\;
 $X$.\textit{SDisj} = $\frac{\displaystyle X.\textit{SConj}}
 {\displaystyle X.\textit{bond}}$\;
$X$.\textit{SNeg} = $\mid$$\mathcal{T}$$\mid$ $-$ $X$.\textit{SDisj}\;

   \Retour{ $\{$$X$, $X$.\textit{SConj},
   $X$.\textit{SDisj},
   $X$.\textit{SNeg},
   $X$.\textit{bond}$\}$ \label{res2}}\;}
           {\Retour{$\emptyset$}\label{res3}\;}
   }}}
\end{algorithm}
\decmargin{0.5em}


\section{Algorithme \textsc{RegenerationMCR} de régénération de $\mathcal{MCR}$} \label{section_regener}
La régénération de l'ensemble $\mathcal{MCR}$ à partir de $\mathcal{RMCR}$ s'effectue grâce à l'algorithme \textsc{RegenerationMCR} dont le pseudo-code est donné par l'algorithme \ref{RCPRegeneration}. Cet algorithme fournit l'ensemble $\mathcal{MCR}$ des motifs corrélés rares munis de leurs supports conjonctifs et de leurs valeurs de la mesure \textit{bond}. L'exemple suivant illustre l'exécution de cet algorithme.

La tâche de régénération s'effectue à travers l'algorithme \textsc{RegenerationMCR} de la manière suivante. D'abord, tous les éléments de la représentation $\mathcal{RMCR}$ seront insérés dans l'ensemble $\mathcal{MCR}$ \textsc{(}\textit{cf.} ligne 4\textsc{)} initialement vide. Par la suite, l'algorithme parcours l'ensemble $\mathcal{MMCR}$ des motifs minimaux et affecte à chaque motif minimal $M$ son fermé $F$ \textsc{(}\textit{cf.} ligne 6\textsc{)}. Puis l'ensemble de motifs compris entre le minimal $M$ et son fermé $F$ est généré \textsc{(}\textit{cf.} ligne 7\textsc{)}. Chaque élément de cet ensemble est un motif corrélé rare et partage le même support conjonctif et la même valeur de \textit{bond} que son fermé $F$ et sera inséré dans l'ensemble $\mathcal{MCR}$ \textsc{(}\textit{cf.} ligne 10\textsc{)}. Lorsque tous les motifs générés sont insérés dans l'ensemble $\mathcal{MCR}$, alors l'algorithme retourne l'ensemble total des motifs corrélés rares $\mathcal{MCR}$
\textsc{(}\textit{cf.} ligne 11\textsc{)}.

\incmargin{1em} \linesnumbered
\begin{algorithm}[h]\label{RCPRegeneration}
 \small{ \small{\caption{\textsc{RegenerationMCR}}
	\SetVline
	\setnlskip{-3pt}
	\Donnees{La représentation concise exacte $\mathcal{RMCR}$ =  $\mathcal{MMCR}$ $\cup$ $\mathcal{MFCR}$.}
\Res{L'ensemble $\mathcal{MCR}$ des motifs corrélés rares munis de leurs valeurs du support conjonctif et de leurs valeurs de la mesure \textit{bond}.}
\Deb{
	 $\mathcal{MCR}$ := $\emptyset$\;
	  \PourCh{\textsc{(}$X \in \mathcal{RMCR}$\textsc{)}}
		  {

 $\mathcal{MCR}$ := $\mathcal{MCR}$ $\cup$ $\{$$X$, $X$.\textit{SConj},
   $X$.\textit{bond}$\}$ \label{majMCR1}\;
  	    }

\PourCh{\textsc{(}$M \in \mathcal{MMCR}$\textsc{)}\label{pour2}}
		  {
       $F$ := 	$\min_{\subseteq}$$\{$$M_{1}$ $\in$  $\mathcal{MFCR}$ $\mid$  $M$  $\subset$ $M_{1}$$\}$ /* $F$ dénote la fermeture du motif minimal corrélé rare $M$, repérée étant le plus petit motif par inclusion ensembliste de la représentation englobant $M$ */ \label{ferme2}\; 	

   \PourCh{\textsc{(}$X$ $\mid$ $M$ $\subset$ $X$ et $X$ $\subset$ $F$\textsc{)}\label{pour3}}
		          {
		          		
  $X$.\textit{SConj} = $F$.\textit{SConj}\;
  $X$.\textit{bond} = $F$.\textit{bond}\;

 $\mathcal{MCR}$ := $\mathcal{MCR}$ $\cup$ $\{$$X$, $X$.\textit{SConj},
   $X$.\textit{bond}$\}$ \label{majMCR2}\;
		          }
             }
 \Retour{$\mathcal{MCR}$}\;\label{resultat}}
}}
\end{algorithm}
\decmargin{1em}


\begin{exemple}
Considérons la représentation concise exacte donnée par l'exemple \ref{ExpRMCR} \textsc{(}\textit{cf.} page \pageref{ExpRMCR}\textsc{)}.
D'abord, l'ensemble $\mathcal{MCR}$ est initialisé par l'algorithme \textsc{RegenerationMCR} à l'ensemble vide. Tous les éléments de $\mathcal{RMCR}$ seront ensuite insérés dans l'ensemble $\mathcal{MCR}$. Ainsi,
$\mathcal{MCR}$ = $\{$\textsc{(}\texttt{D}, 1, $\frac{1}{1}$\textsc{)},
\textsc{(}\texttt{AB}, 2, $\frac{2}{5}$\textsc{)},
\textsc{(}\texttt{AD}, 1, $\frac{1}{3}$\textsc{)},
\textsc{(}\texttt{AE}, 2, $\frac{2}{5}$\textsc{)},
\textsc{(}\texttt{CD}, 1, $\frac{1}{4}$\textsc{)},
\textsc{(}\texttt{ACD}, 1, $\frac{1}{4}$\textsc{)},
\textsc{(}\texttt{ABCE}, 2, $\frac{2}{5}$\textsc{)}$\}$.
Par la suite, nous générons les motifs \texttt{ABE} et \texttt{ABC} compris entre le minimal \textsc{(}\texttt{AB}, 2, $\frac{2}{5}$\textsc{)} et son fermé \textsc{(}\texttt{ABCE}, 2, $\frac{2}{5}$\textsc{)} et le motif \texttt{ACE} compris entre le minimal \textsc{(}\texttt{AE}, 2, $\frac{2}{5}$\textsc{)} et son fermé \textsc{(}\texttt{ABCE}, 2, $\frac{2}{5}$\textsc{)}. Les motifs \texttt{ABE}, \texttt{ABC} et \texttt{ACE} générés seront alors insérés dans l'ensemble $\mathcal{MCR}$. Ce dernier englobe, ainsi, tous les motifs corrélés rares.
$\mathcal{MCR}$ = $\{$\textsc{(}\texttt{D}, 1, $\frac{1}{1}$\textsc{)},
\textsc{(}\texttt{AB}, 2, $\frac{2}{5}$\textsc{)},
\textsc{(}\texttt{AD}, 1, $\frac{1}{3}$\textsc{)},
\textsc{(}\texttt{AE}, 2, $\frac{2}{5}$\textsc{)},
\textsc{(}\texttt{CD}, 1, $\frac{1}{4}$\textsc{)},
\textsc{(}\texttt{ABC}, 2, $\frac{2}{5}$\textsc{)},
\textsc{(}\texttt{ABE}, 2, $\frac{2}{5}$\textsc{)},
\textsc{(}\texttt{ACD}, 1, $\frac{1}{4}$\textsc{)},
\textsc{(}\texttt{ACE}, 2, $\frac{2}{5}$\textsc{)},
\textsc{(}\texttt{ABCE}, 2, $\frac{2}{5}$\textsc{)}$\}$.
\end{exemple}


\section{{\'E}valuation expérimentale de la représentation $\mathcal{RMCR}$} \label{section_XP}
Notre objectif principal, dans cette section, est de prouver expérimentalement le taux de compacité de la représentation $\mathcal{RMCR}$. Les différentes expérimentations réalisées ont été menées sur une machine munie d'un processeur Intel Dual Core $E5400$, ayant une fréquence de $2,7$GHz avec $4$Go de mémoire vive, tournant sur une plateforme Linux Ubuntu $10.04$. Les expérimentations ont été réalisées sur différentes bases de test benchmark denses et éparses $^{\textsc{(}}$\footnote{Disponibles à l'adresse suivante : \textsl{http://fimi.cs.helsinki.fi/data}.}$^{\textsc{)}}$.


\begin{figure}[h]
\parbox{16cm}{\hspace{-0.5cm}
 \parbox{4.6cm}{\includegraphics[scale = 0.65]{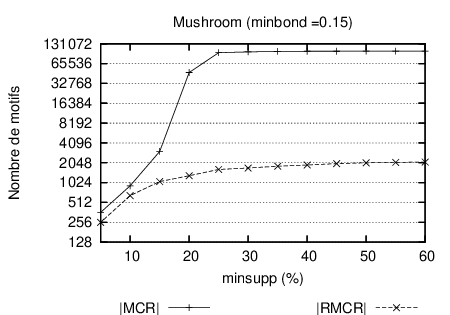}}
 \parbox{4.6cm}{\includegraphics[scale = 0.65]{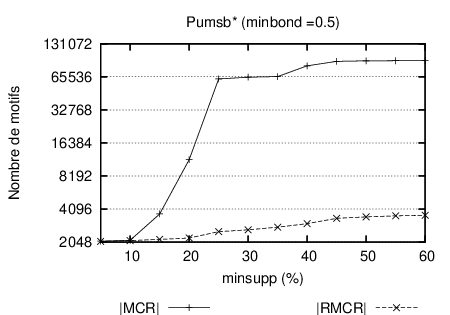}}
 \parbox{4.6cm}{\includegraphics[scale = 0.65]{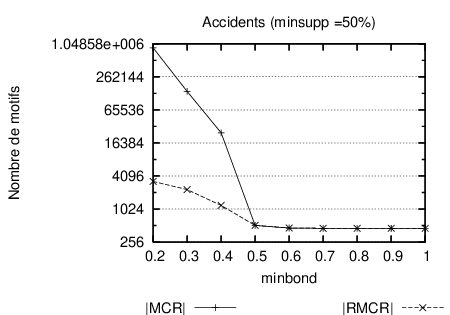}}}
 \caption{Variation des cardinalités de la représentation $\mathcal{RMCR}$ \textit{versus} celles de l'ensemble $\mathcal{MCR}$ en fonction de \textit{minsupp} et de \textit{minbond}.}\label{Fct_minsupp}
\end{figure}

Les résultats expérimentaux les plus représentatifs sont donnés par la figure \ref{Fct_minsupp}. Nous constatons, d'après ces résultats, que les taux de réduction obtenus pour la représentation proposée et pour différents seuils \textit{minsupp} et \textit{minbond} sont intéressants. Par ailleurs, la représentation $\mathcal{RMCR}$ est prouvée être une couverture parfaite de l'ensemble $\mathcal{MCR}$ dans le sens que sa taille ne dépasse jamais celle de ce dernier. Considérons par exemple, la base \textsc{Mushroom} pour \textit{minsupp} = \textit{35}$\%$ et \textit{minbond} = \textit{0,15} : $|\mathcal{RMCR}|$ = \textit{1 810}, et $|\mathcal{MCR}|$ = \textit{100 156}. Le taux de compacité dans ce cas est de \textit{98}$\%$. Ces résultats sont obtenus grâce à la propriété de non-injectivité de l'opérateur de fermeture $f_{bond}$. En effet, cet opérateur permet de regrouper les motifs ayant les mêmes propriétés dans une même classe d'équivalence. Ceci permet ainsi d'éviter la redondance des éléments maintenus. Nous avons, par exemple, pour la base \textsc{Mushroom} : $|\mathcal{MMCR}|$ = \textit{1 412} et $|\mathcal{MFCR}|$ = \textit{652}. Puisque la représentation $\mathcal{RMCR}$ correspond à l'union sans redondance des ensembles $\mathcal{MMCR}$ et $\mathcal{MFCR}$, nous avons toujours $|\mathcal{RMCR}|$ $\leq$ $|\mathcal{MMCR}|$ + $|\mathcal{MFCR}|$.

Dans ce qui suit, nous proposons une application de $\mathcal{RMCR}$ dans le cadre de la détection d'intrusions dans les réseaux informatiques.
\section{Application de la représentation $\mathcal{RMCR}$ dans la détection d'intrusions} \label{section_IDS}
Nous présentons dans cette section, l'application de la représentation $\mathcal{RMCR}$ dans un processus de classification basé sur les règles d'association corrélées rares. En effet, les ensembles de motifs $\mathcal{MMCR}$ et $\mathcal{MFCR}$, composant la représentation $\mathcal{RMCR}$, seront incorporés dans la dérivation des règles d'association génériques corrélées rares de la forme $Gen$ $\Rightarrow$ $Ferm$é $\setminus$ $Gen$, avec $Gen$ $\in$ $\mathcal{MMCR}$ et $Ferm$é $\in$ $\mathcal{MFCR}$ $^\textsc{(}$\footnote{Par ``générique'', nous entendons que ces règles sont à prémisse minimale et à conclusion maximale, selon la relation d'inclusion ensembliste.}$^{\textsc{)}}$.

Ensuite, à partir des règles génériques obtenues, nous extrayons les règles de classification. En effet, les règles génériques obtenues seront filtrées afin de ne garder que les règles génériques ayant le libellé de la classe d'attaque dans la partie conclusion. Ces règles seront alors communiquées au classifieur que nous avons conçu. Ce dernier permet d'élaborer la tâche de classification et retourne le taux de détection pour chaque classe d'attaque. Nous présentons dans la suite l'évaluation expérimentale de la classification basée sur les règles corrélées rares
pour la base de données \textsc{KDD 99} $^\textsc{(}$\footnote{La base \textsc{KDD 99} est disponible à l'adresse suivante : \textsl{http://kdd.ics.uci.edu/databases/kddcup99/kddcup99.html}.}$^\textsc{)}$.

\subsection{Description de la base \textsc{KDD 99}} \label{DataDescrip}
Chaque objet de la base \textsc{KDD 99} représente une connexion du flot de données. Une connexion est ainsi étiquetée comme \textit{Normale} ou \textit{Attaque}. La base \textsc{KDD 99} décrit 38 catégories d'attaques réparties en quatre classes d'attaques, à savoir \textsc{Dos}, \textsc{Probe}, \textsc{R2L} et \textsc{U2R}, et une classe \textsc{Normale}. Cette base contient 4 940 190 objets dans la base d'apprentissage et chaque objet est caractérisé par 41 attributs. Nous considérons, dans ce travail, 10$\%$ de l'ensemble d'apprentissage dans la phase de construction du classifieur, contenant ainsi 494 019 objets. L'ensemble d'apprentissage contient 79,20$\%$ \textsc{(}respectivement, 0,83$\%$, 0,22$\%$ et 0,10$\%$\textsc{)} d'attaques \textsc{Dos} \textsc{(}respectivement, \textsc{Probe}, \textsc{R2L} et \textsc{U2R}\textsc{)}, et le reste, \textit{c.-à.-d.} 19,65$\%$, concerne des connexions étiquetées \textit{Normale}.

\subsection{Discussion des résultats obtenus} \label{RecapXp}
Les résultats expérimentaux obtenus sont donnés par la table \ref{TabXpKDD3}, avec ``\textsc{RA}s'' et ``\textsc{TD}'' les abréviations respectives de ``Règles d'Association'' et ``Taux de Détection'',
et \textit{minconf} dénote le seuil minimal de la mesure \textit{confiance} \citep{Agra94}.
Nous entendons aussi par ``Phase de construction'' l'étape de l'extraction de la représentation $\mathcal{RMCR}$
tandis que par ``Phase de classification'', nous entendons l'étape de dérivation des règles de classification à partir de la représentation $\mathcal{RMCR}$ et leur application dans la détection d'intrusions.

Nous constatons que les taux  de détection les plus intéressants sont achevés pour les classes d'attaques
\textsc{Normale} et \textsc{Dos}. En effet, ceci est expliqué par la taille élevée en nombre de connections de ces deux classes d'attaques. Ceci confirme que notre approche proposée dans ce travail présente de meilleures performances pour des bases volumineuses. Nous remarquons aussi que ce taux de détection varie d'une classe d'attaque à une autre. Par exemple, pour la classe \textsc{U2R}, ce taux est relativement faible par rapport aux autres classes d'attaques.

Nous concluons aussi, d'après les résultats de la table \ref{TabXpKDD3}, que les coûts de calcul varient d'une classe d'attaque à une autre. Toutefois, pour les différentes classes d'attaques considérées, la phase de construction est plus coûteuse en temps d'exécution que la phase de classification. Ceci est justifié par le fait que l'étape de construction englobe l'extraction de la représentation concise $\mathcal{RMCR}$, or cette opération est NP-difficile  \citep{boley2009} étant donnée la complexité liée à la localisation des deux bordures associées aux contraintes de corrélation et de rareté.

\font\xmplbx = cmbx7.5 scaled \magstephalf
\begin{table}[htbp]
\vspace{-0.1cm}
\hspace{-2.2cm}
\footnotesize{
\begin{tabular}{|c||r|r|r||r|r|r||r||r|r|} \hline
\textbf{Classe}& \textbf{\textit{minsupp}}&\textbf{\textit{minbond}} &\textbf{\textit{minconf}} &\textbf{$\#$ \textsc{RA}s}&\textbf{$\#$ \textsc{RA}s} & \textbf{$\#$ \textsc{RA}s}&\textbf{\textsc{TD}}&\multicolumn{2}{c|}{\textbf{Temps CPU \textsc{(}en secondes\textsc{)}}}\\
   \textbf{de l'attaque} & \textbf{\textsc{(}$\%$\textsc{)}} &  & &\textbf{génériques}&\textbf{génériques}& \textbf{génériques}&\textbf{\textsc{(}$\%$\textsc{)}}&\textbf{Phase de}& \textbf{Phase de} \\
   & & & &\textbf{exactes} &\textbf{approximatives}&\textbf{de classification}&&  \textbf{construction}&  \textbf{classification}\\  \hline
 \hline \textsc{Dos}   & {\xmplbx80}  & {\xmplbx0,95}& {\xmplbx0,90} & {\xmplbx4} & {\xmplbx31}  &{\xmplbx17}&{\xmplbx98,68}&{\xmplbx120}&{\xmplbx1}\\
\hline \textsc{Probe} & {\xmplbx60}  &{\xmplbx0,70} &{\xmplbx0,90} & {\xmplbx232} & {\xmplbx561}  &{\xmplbx15}&{\xmplbx70,69}&{\xmplbx55}&{\xmplbx1}\\
\hline \textsc{R2L}   & {\xmplbx80}  &{\xmplbx0,90} &{\xmplbx0,70} & {\xmplbx2}& {\xmplbx368} &{\xmplbx1}&{\xmplbx81,52}&{\xmplbx1 729}&{\xmplbx1}\\
\hline \textsc{U2R}   & {\xmplbx60}  &{\xmplbx0,75} &{\xmplbx0,75} & {\xmplbx106}& {\xmplbx3}  &{\xmplbx5}&{\xmplbx38,46}&{\xmplbx32}&{\xmplbx1}\\
\hline \textsc{Normale}& {\xmplbx85}  &{\xmplbx0,95} &{\xmplbx0,95} & {\xmplbx0} & {\xmplbx10}  &{\xmplbx3}&{\xmplbx100,00}&{\xmplbx393}&{\xmplbx15}\\
\hline
\end{tabular}}
\caption{\'Evaluation des règles d'association corrélées rares pour la base \textsc{KDD 99}.}
\label{TabXpKDD3}
\end{table}

La table \ref{TabCmp} compare les résultats obtenus par notre approche, basée sur les règles d'association
corrélées rares, à ceux offerts par les approches basées respectivement sur les arbres de décisions et les réseaux bayésiens \citep{NahlaSac2004}. Il est à noter que le choix de ces approches pour ce comparer avec est argumenté par le fait que celle utilisant les arbres de décisions est aussi basée sur les règles d'association. Par ailleurs, l'apprentissage est supervisé dans les différentes approches comparées. Les résultats obtenus prouvent que notre approche offre dans différentes situations de meilleures performances que les autres approches. En effet, elle est la meilleure pour les classes d'attaques \textsc{Dos}, \textsc{R2L} et \textsc{U2R}. Bien que aussi meilleurs pour la classe \textsc{Normale}, les résultats obtenus sont très proches de ceux obtenus avec les arbres de décision. Les réseaux bayésiens présentent de meilleurs taux de détection uniquement pour la classe \textsc{Probe}. Ainsi, l'application des règles corrélées rares offre une solution intéressante dans le contexte de la détection d'intrusions.

\font\xmplbx = cmbx7.5 scaled \magstephalf
\begin{table}[htbp]
\begin{center}
\footnotesize{\begin{tabular}{|c||r||r|r|} \hline
 \textbf{Classe d'attaque}   &\textbf{\textsc{RA}s corrélées rares}&\textbf{Arbres de décision}&\textbf{Réseaux bayésiens}  \\
 \hline  \hline \textsc{Dos}   & {\xmplbx98,68} &97,24& 96,65\\
\hline \textsc{Probe} & 70,69 & 77,92&{\xmplbx88,33}\\
\hline \textsc{R2L}   & {\xmplbx81,52} &0,52&8,66\\
\hline \textsc{U2R}   & {\xmplbx38,46} &13,60&11,84\\
\hline \textsc{Normale}& {\xmplbx100,00}&99,50&97,68\\
\hline
\end{tabular}}
\caption{Comparaison des taux de détection obtenus pour les règles corrélées rares \textit{versus} les approches de l'état de l'art.}
\label{TabCmp}
\end{center}
\vspace{-1cm}
\end{table}
\section{Conclusion et perspectives}\label{section_cl}
Dans ce papier, nous avons proposé l'algorithme \textsc{RcprMiner} d'extraction de la représentation concise exacte $\mathcal{RMCR}$ de l'ensemble $\mathcal{MCR}$ des motifs corrélés rares. Nous avons introduit également l'algorithme \textsc{EstMCR} d'interrogation de cette représentation ainsi que l'algorithme \textsc{RegenerationMCR} de dérivation de l'ensemble $\mathcal{MCR}$ à partir de $\mathcal{RMCR}$. Nous avons démontré expérimentalement le taux de réduction intéressant offert par cette représentation. L'efficacité de la classification, basée sur les règles d'association corrélées rares, a été aussi prouvée dans le cadre de la détection d'intrusions.

Les perspectives de travaux futurs concernent : \textsc{(}$i$\textsc{)} La comparaison détaillée des performances d'un algorithme d'extraction de $\mathcal{MCR}$, directement à partir d'une base de transactions, à celles de \textsc{RcprMiner} suivi par \textsc{RegenerationMCR} pour dériver l'ensemble total des motifs corrélés rares à partir de $\mathcal{RMCR}$. Ceci permettra de cerner aussi les situations où le recours à la représentation $\mathcal{RMCR}$ est aussi nécessaire non seulement pour réduire la taille des connaissances extraites mais aussi pour rendre possible la fouille des motifs corrélés rares. \textsc{(}$ii$\textsc{)} L'extraction, à partir de $\mathcal{RMCR}$, de formes généralisées de règles d'association présentant des conjonctions, des disjonctions, et des négations d'items en prémisse ou en conclusion ainsi que leur application dans des contextes réels. \textsc{(}$iii$\textsc{)} L'extension de l'approche proposée pour d'autres mesures de corrélation \citep{Kim-pkdd2011,Omie03,borgelt,surana2010,Xiong06hypercliquepattern} en se basant sur l'étude de leurs propriétés respectives. 
{\small\bibliographystyle{rnti}
\bibliography{Refs}}
\end{document}